\documentstyle [12pt] {article}
\begin{document}
\begin{titlepage}
\begin{center}
\vspace{2cm}
\LARGE
Constraints on Models of Galaxy Formation from the Evolution of Damped
Ly$\alpha$
Absorption Systems\\
\vspace{1cm}
\large
Guinevere Kauffmann  and St\'{e}phane Charlot \\
\vspace{0.5cm}
{\em Dept. of Astronomy, University of California, Berkeley, CA, 94720}\\
\vspace{0.8cm}
\end{center}
\large
\begin {abstract}
\baselineskip 0.8 cm
There is accumulating observational evidence suggesting that damped Ly$\alpha$
absorption systems
systems are the progenitors of
present-day spiral galaxies. We use the observed properties of these systems to
place
constraints on the history of star formation in galactic disks, and on
cosmological
theories of structure formation in the universe.
We show that  the observed increase in $\Omega_{HI}$
contributed by damped Ly$\alpha$ systems at high redshift
implies that star formation must have been considerably less efficient in the
past.
We construct a model in which gas is converted into stars with an efficiency
that
increases with time, and we show that this model can reproduce most of the
observed
properties of damped Ly$\alpha$ systems, including the observed distribution of
column densities.
We also show that the data can constrain cosmological models in which structure
forms at late epochs.
A mixed dark matter (MDM) model with $\Omega_{\nu}=0.3$
is unable to reproduce the mass densities of cold
gas seen at high redshift, even in the absence of {\em any} star formation.
We show that at redshifts greater than 3, this model predicts that
the total baryonic mass contained
in dark matter halos with circular velocities
$V_c > 35$ km s$^{-1}$ is less than the observed mass of HI in damped systems.
At these redshifts, the photo-ionizing background
would prevent gas from dissipating and collapsing to form high column density
systems in
halos smaller than 35 km s$^{-1}$. MDM models are thus ruled out by the
observations.
\end {abstract}
\normalsize
{\em Subject headings:} galaxies:formation, galaxies:evolution, quasars:
absorption lines,
cosmology:theory
\end {titlepage}

\section {Introduction}
\baselineskip 0.8 cm
Absorption-line features in the spectra of distant quasars arising from
intervening
damped Ly$\alpha$ systems
are presently yielding extremely interesting information about the
evolution of the gaseous content of the universe. Surveys carried out at
optical
wavelengths indicate that the
damped Ly$\alpha$ absorbers contribute a cosmological mass density roughly
comparable
to the density of luminous matter in present-day spiral galaxies (Wolfe et al.
1986).
More recently, Lanzetta et al. (1993) have
analyzed a set of International Ultraviolet Explorer (IUE) spectra in order to
extend the
search for damped Ly$\alpha$ absorption systems to lower redshifts.
Taken together, these surveys indicate that the
cosmological mass density of neutral gas associated with damped Ly$\alpha$
absorption
decreases significantly from $z=3.5$ to $z=0.008$, and this evolution results
from a steady decrease in the incidence of high column density absorption
systems with
decreasing redshift.

The damped Ly$\alpha$ absorbers are prime candidates for the
progenitors of spiral galaxies. Detailed absorption-line modelling
of individual damped Ly$\alpha$ absorption systems indicates a resemblance to
extended
HI disks
(Wolfe et al. 1993).
Further evidence for the disk hypothesis has also come using a variety of other
observational
techniques. VLBI investigation of
the $z=2.04$ 21cm line in the quasar PKS0458-02 reveals that the absorber
extends at least
$8 h^{-1}$ kpc transverse to the line of sight (Briggs et al. 1989).
In addition, Wolfe et al. (1992) have acquired deep CCD images of the field
surrounding
the QSO H0836+113 and find a Ly$\alpha$ emitter corresponding to the z=2.466
damped
Ly$\alpha$ line with
linear dimensions $24 \times 11 h^{-2}$ kpc$^{2}$, comparable to the size of a
present-day
disk galaxy.

If the damped Ly$\alpha$ systems are indeed the progenitors of galactic disks,
then the
observed evolution of these systems with redshift places important constraints
on
theoretical models of galaxy formation. Most currently popular cosmological
models,
including cold dark matter (CDM) models and variations thereof, predict that
structure
forms in a hierarchical fashion, with small objects merging to form larger and
larger
structures. Gas, which is trapped and heated within a collapsing dark matter
halo, may
then cool and collapse to form a rotationally supported disk at the centre of
the halo.
This picture has been investigated in a number of papers using analytic
techniques based
on the Press-Schechter formalism (White \& Frenk 1991; Lacey \& Silk 1991;
 Kauffmann et al. 1993;
Lacey et al. 1993; Cole et al. 1994). It has also been studied using numerical
techniques
such as smoothed particle hydrodynamics (SPH) (Katz \& Gunn 1991; Katz,
Hernquist \& Weinberg
1992; Navarro \& White 1994; Evrard, Summers \& Davis 1993).

In this paper, we use the most recent data of Lanzetta et al. (1993) on the
statistics of
damped Ly$\alpha$ absorption systems
to constrain cosmological models
and the history of star formation in galaxies. We make use of the
analytic formalism developed by Kauffmann \& White (1993) and Kauffmann et al.
(1993) to
keep track of the cooling of gas and formation of stars
in an evolving distribution of dark matter
halos in a hierarchical universe.

In section 3, we show that the rapid increase in the
cosmological mass density of neutral gas at high redshifts cannot be
accommodated in {\em any}
model where stars form with the same efficiency at all epochs.
In most studies of galaxy formation in a hierarchical universe,
stars are assumed to form at a rate proportional to the amount of gas which
cools
at any given time. This leads to neutral gas mass densities that remain
constant or even
decrease at high redshift.
We also show that a mixed dark matter (MDM) model with $\Omega_{\nu}=0.3$ is
not able to
account for the large
amounts of HI gas observed at high redshift in damped Ly$\alpha$ systems,
simply because structure
forms too late in this model.
In section 4, we construct a model in which the efficiency of star formation in
the universe
decreases as a power of the redshift $z$,
and we assume that gas cooling within a dark matter halo forms an exponential
disk
with a scale length proportional to the virial radius of the halo.
We find that our model is in good agreement with the available observations.
Finally, we show that our empirically-derived
star formation law leads to disk galaxies with spectral
properties similar to those of present day spiral galaxies.

\section {A Brief Description of the Galaxy Formation Model}
The semianalytic model of galaxy formation that we use
has been described in detail in a recent paper by Kauffmann, White \&
Guiderdoni (1993, hereafter KWG). We will
only present a brief summary of the model here.

We use an extension of the Press-Schechter theory due to Bower (1991) and Bond
et al. (1991)
to model the evolution of dark matter halos in a hierarchical universe.
We construct Monte Carlo realizations of the merging history of a present-day
dark matter
halo of  given mass.
This allows us to trace the merging path of every halo mass element through all
the
progenitor halos from which the present object formed.

Let us now consider a set of dark matter halos present in the universe at some
redshift z.
For simplicity, we model our halos as truncated, singular, isothermal spheres.
We assume that when a halo forms, gas is shock-heated to the virial temperature
of the halo
and relaxes to a distribution that exactly parallels that of the dark matter.
Gas is then able to cool and collapse to form a galaxy at the centre of the
halo.
In the terminology adopted by KWG, this galaxy is referred to as the {\em
central}
galaxy of the halo.

The rate at which cold gas settles onto the central galaxy of a halo of
circular velocity $V_{c}$ is given by
min$(\dot{M}_{inf},\dot{M}_{cool})$. The infall rate $\dot{M}_{inf}$
onto the halo is given by
\begin{equation} \dot{M}_{inf}(V_{c},z)=0.1 f_{g} \Omega_{b} V_{c}^{3}
G^{-1},\end {equation}
where $\Omega_{b}$ is the fraction of the critical density in the form of
baryons and $f_g$ is the fraction of the baryons in the form of gas.
In this paper, we adopt $\Omega_{b}=0.06$, which is the upper limit of values
allowed by
Big Bang nucleosynthesis if $H_0= 50$ km s$^{-1}$ Mpc$^{-1}$ (Olive et al.
1990).
The cooling rate $\dot{M}_{cool}$ can be expressed as a simple inflow
equation,
\begin {equation} \dot{M}_{cool}(V_{c},z)= 4 \pi \rho_{g}(r_{cool})
r^{2}_{cool} \frac{dr_{cool}}{dt}, \end{equation}
where $r_{cool}$ is the radius at which the cooling time in the halo is equal
to the age of the universe.

It is not easy to predict exactly at what rate this cold gas will turn into
stars.
KWG choose
a star formation law given by the simple equation,
\begin {equation} \dot{M_*}= \alpha \frac {M_{cold}}{t_{dyn}} \end {equation}
where $M_{cold}$ is the total mass of cold gas, $t_{dyn}$ is a characteristic
dynamical timescale for the galaxy, and $\alpha$ is an adjustable parameter.
Similar star formation laws have also been adopted by other authors
(White \& Frenk 1991; Cole et al. 1994).
For a standard Scalo (1986) IMF, the number of supernovae expected per solar
 mass of stars formed
is $\eta_{SN} \sim 4 \times 10^{-3} M_{\odot}^{-1}$. The kinetic energy of
the ejecta from each supernova, $E_{SN}$, is about 10$^{51}$ ergs. If
a fraction $\epsilon$ of this energy is used to reheat cold gas to
the virial temperature of the halo, the amount of cold gas lost
to the interstellar medium in time $\Delta t$ can be estimated as,
\begin{equation} \Delta M_{reheat}=\epsilon \frac {4}{5}\frac {\dot{M_*}
\eta_{SN}
E_{SN}}{V_{c}^{2}} \Delta t. \end{equation}
The parameters $\alpha$ and $\epsilon$ together control the luminosity and cold
gas content
of a galaxy. In KWG, these parameters were fixed so that the central galaxy of
a
$V_{c}=220$ km s$^{-1}$ halo at z=0 had a luminosity and gas content equal to
that of
the Milky Way.

Let us now consider this set of halos at some later redshift, $z_{0} < z$.
Our Monte Carlo merging histories tell us which halos
present at redshift $z$ have merged with each other to form a larger object.
The central galaxy of this
new halo is identified with the central galaxy of its largest predecessor. The
galaxies
that have formed inside the smaller predecessors become {\em satellites}. These
satellites
merge with the central galaxy at a rate governed by the dynamical
friction timescale (see KWG for more details).

\section {Constraints from the Evolution of $\Omega_{HI}$}
In this section, we place constraints on cosmological models and on the star
formation
history of galaxies using the observed evolution of the HI mass density
contributed by
damped Ly$\alpha$ systems. We consider three cosmological models:
\begin {enumerate}
\item A b=1.5 CDM model. This is the lowest value of $b$ allowed by the
observed abundance of rich clusters (White et al. 1993). A COBE normalized
model would
have b=1.
\item A b=2.5 CDM model.
\item A MDM model with $\Omega_{\nu}=0.3$ normalized to COBE.
This value of the neutrino mass density has
been shown to provide the best fit to the observed large scale structure and
streaming motions of galaxies (Klypin et al. 1993; Pogosyan \& Starobinsky
1993).
\end {enumerate}
In all three cases, we assume $\Omega=1$ and $H_0= 50$ km s$^{-1}$ Mpc$^{-1}$.
In our formalism, the influence of the choice of cosmological model is
reflected by the
way the mass distribution of dark matter halos changes
as a function of redshift. Structure forms progressively later as we proceed
from
model 1 to model
3 in the list above. In the MDM model, Galaxy-sized halos only become abundant
at redshifts
less than 1.

The equations governing cooling and star
formation within the halos have already been described in section 2. Note that
equation (3)
implies that stars always form at a rate proportional to
the mass of cold gas in the halo. In figure 1,
we plot the total mass density of cold gas contained in halos with circular
velocities
between 50 and 300 km s$^{-1}$, corresponding to the circular velocities of
present-day
spiral galaxies.
As can be seen, the predicted $\Omega_{HI}$ matches the
observations at low redshift in all three models. This is not unexpected
because we have tuned the parameters $\alpha$ and $\epsilon$ in equations 3 and
4
to match the gas mass of a Milky Way-type galaxy at the present day. However,
$\Omega_{HI}$ does not increase with redshift in any of these models as in the
data.
It either remains flat or, in the case of the MDM model, drops off at high
redshift.
It is clear than in
order to reproduce the observed increase in the HI mass density, stars must
form less
efficiently at high redshift.

We now show that structure formation occurs too late in the MDM model to
account for the
large mass of high column-density neutral hydrogen observed at large redshifts,
even if we neglect star formation completely.
The short-dashed/dotted  line in figure 1 shows the
mass density contributed by {\em all} baryons in
halos with circular velocities between 50 and 300 km s$^{-1}$. Recall that we
have chosen
$\Omega_{b}=0.06$. This curve falls below the observational data at redshifts
greater than 2.
The hypothesis that damped Ly$\alpha$
systems are the progenitors of spiral galaxies is simply not viable for an MDM
cosmology.
We can make an even stronger case against MDM by considering the mass density
contributed
by baryons in halos with $V_c > 35$ km s$^{-1}$ (the long-dashed/dotted line in
figure 1).
This curve falls below the observations at $z>3$. Rees (1986) has shown that an
external
UV background of an intensity 10$^{-21}$ erg cm$^{-2}$ s$^{-1}$ Hz$^{-1}$
sr$^{-1}$,
sufficient to satisfy the Gunn-Peterson constraints on the ionization of the
diffuse
intergalactic medium and explain the proximity
effect at redshifts $z \sim 2-3$ (Lu et al. 1991),
would cause gas to be stably confined in halos with $V_c < 35$ km s$^{-1}$.
Gas would then be unable to dissipate and collapse to form a system with high
column-density.
MDM models thus appear to be ruled out even if damped Ly$\alpha$ systems have
no connection with present-day disk galaxies. This is not true of either of the
two CDM
models we have considered.

\section {A Fit to the Observational Data}
In this section, we present a model which can fit the observations.
We modify the star formation law in equation 3 to allow for
a star formation efficiency which decreases with redshift, and we focus on the
b=1.5 CDM
model. We now write
\begin {equation} \dot M_* = \alpha_{0} (z_{i}-z)^{n} M_{cold}/t_{dyn} \end
{equation}
where $z_{i}$ is the redshift at which star formation ``switches on''.
The heavy solid line in figure 1 shows the evolution of
$\Omega_{HI}$ for a b=1.5 CDM model and the
parameters $\alpha_{0}=0.002$, $z_{i}=4.0$ and $n=2$.  We are now able to
reproduce the
observed increase in $\Omega_{HI}$ at high redshift.
These parameters also leads to a
``Milky Way'' gas mass and luminosity that matches observations.

Next we investigate whether we can reproduce the observed distribution of
damped Ly$\alpha$
absorber column densities. We assume for simplicity that gas cooling in a halo
collapses
to form a planar disk with a radial HI profile $N_{\perp}(r)$ when viewed face
on.
We also assume that this profile has exponential form: $N_{\perp}(r)= N_{\perp
0} \exp(- r/a)$,
 where $N_{\perp 0} =  M_{gas}/( 2 \pi a^2)$ is the central column
density of the disk.

Bosma (1981) has made a detailed study of the radial distribution of HI gas in
spiral
galaxies. He finds that N(HI) typically decreases from a few times $10^{21}$
cm$^{-2}$
at the centre of the disk to $10^{20}$ cm$^{-2}$ at about 1.5 Holmberg radii.
The
Holmberg radius for a typical $L_{*}$ galaxy is about 15-18 kpc. This implies
an
HI disk scale length of about 8 kpc. We have adopted a simple scaling
relation for the scale length of disks forming inside a halo of circular
velocity
$V_{c}$ at redshift $z$:
\begin {equation} a = 8 \left( \frac {V_{c}} {220 km s^{-1}} \right)
(1+z)^{-1.5} kpc. \end {equation}
This is motivated by the relationship between the virial radius of a dark
matter halo and
its circular velocity (White \& Frenk 1991)
\begin {equation} r_{vir}=0.1 H_{0}^{-1} (1+z)^{-1.5} V_{c}. \end {equation}

Following Fall \& Pei (1993), the column density distribution at redshift z may
be written
\begin {equation} f_{t}(N,z)= \int  \left ( \frac{\pi c \mu_m a^2} {2 H_0  N}
\right )
  n(V_c) [ 1 - 2 \ln (\mu_m N / N_{\perp 0}) ] dV_c, \end {equation}
where $n(V_{c})$ is the number density of halos with circular velocity in the
range
$(V_c, V_c+dV_c)$ and
$\mu_m=1$ for $N \le N_{\perp 0}$ and $ \mu_{m}= N_{\perp 0}/N$ for $N \ge
N_{\perp 0}$.
 The resulting column density distributions for 3 different redshift bins are
shown
in figure 2. We see that our exponential disk model is able to provide a
reasonable
match to the observations, including the increase in the number
of high column density systems with redshift.

In figure 3a, we show the star formation rate as a function of redshift for a
Milky Way-type
galaxy. The star formation rate increases rapidly, peaking at 8 $M_{\odot}$
yr$^{-1}$
at a redshift 1 and then declining to about 2 $M_{\odot}$ yr$^{-1}$ by the
present day as
all the cold gas is used up in the galaxy. We have used the spectral synthesis
models of
Bruzual \& Charlot (1993)  and a Scalo IMF to compute the
spectrum of a galaxy with this star formation history
at the present day. In figure 3b we compare the resulting spectrum with the
observed spectra of typical Sb and Sc galaxies. The spectra are taken from
Kennicutt (1992).
The model spectrum has a 4000 \AA \hspace{2mm} break and optical colours
intermediate between those
of the two galaxies. We therefore conclude that our
empirically-motivated star formation law (equation 6) can produce disks with
spectral
energy distributions consistent with those of present-day galaxies like the
Milky
Way.

\section {Discussion and Conclusions}
In this paper, we have assumed that damped Ly$\alpha$ systems are the
progenitors of
present-day spiral galaxies and we have used their observed properties to place
constraints on cosmological models and the history of star formation in disks.
We have shown that the observed evolution of $\Omega_{HI}$ contributed by these
systems
implies that stars must have formed
considerably less efficiently in the past. MDM models are unable to reproduce
the mass
of cold gas seen at high redshift, even in the absence of star formation.
This model predicts that at redshifts greater than 3,
most of the mass in the universe is in halos too small
to permit photo-ionized gas to dissipate and collapse to form high
column-density
systems.

We present an empirically-motivated star formation law that can reproduce
most of the observed properties of damped Ly$\alpha$ systems in a b=1.5
CDM cosmology. Using a simple model for an absorber as a planar exponential
disk,
we are able to match the evolution of $\Omega_{HI}$ and the
distribution of column densities  as a function of redshift.
For b=1.5 CDM, we require that the star formation efficiency decreases in
proportion to the
square of the redshift and that no stars form in disks at
redshifts greater than 4. This upper limit to the redshift of star formation
would be
relaxed in a model
where structure forms at an earlier epoch, for example in low-density
models with $\Omega \sim 0.2$ or in models with  higher normalization (lower
$b$).

It is not clear what physical mechanism would cause star formation to be less
efficient
at high redshift. Perhaps certain processes such as satellite accretion occur
more
frequently in the past and are able to thicken the disk and stabilize it
against perturbations.
These are issues which need to be explored more deeply if the connection
between
galaxy formation and the statistics of damped Ly$\alpha$ absorption systems
is to be understood.

\\

\vspace {0.8cm}

\large
{\bf Acknowledgments}\\
\normalsize
We thank Joe Silk and Simon White for helpful discussions.

\pagebreak
\Large
\begin {center} {\bf References} \\
\end {center}
\vspace {1.5cm}
\normalsize
\parindent -7mm
\parskip 3mm

Bond, J.R., Cole, S., Efstathiou, G. \& Kaiser, N. 1991, ApJ, 379, 440

Bosma, A. 1981, AJ, 86, 1825

Bower, R. 1991, MNRAS, 248, 332

Briggs, F.H., Wolfe, A.M., Liszt, H.S., Davis, M.M. \& Turner, K.L. 1989,
ApJ, 341, 650

Bruzual, G. \& Charlot, S. 1993, ApJ, 405, 538

Cole, S., Aragon-Salamanca, A., Frenk, C.S., Navarro, J.F. \& Zepf, S.E.
1994, MNRAS, in press

Evrard, A.E., Summers, F. \& Davis, M. 1993, preprint

Fall, S.M. \& Pei, Y.C. 1993, ApJ, 402, 479

Katz, N. \& Gunn, J.E. 1991, ApJ, 377, 365

Katz, N., Hernquist, L. \& Weinberg, D.H. 1992, ApJ, 399, L109

Kauffmann, G. \& White, S.D.M. 1993, MNRAS, 261, 921

Kauffmann, G., White, S.D.M. \& Guiderdoni, B. 1993, MNRAS, 264, 201 (KWG)

Kennicutt, R. 1992, ApJS, 79, 255

Klypin, A., Holtzman, J., Primack, J. \& Regos, E. 1993, ApJ, 416, 1

Lacey, C. \& Silk, J. 1991, ApJ, 381, 14

Lacey, C., Guiderdoni, B., Rocca-Volmerange, B. \& Silk, J. 1993, ApJ,
402, 15

Lanzetta, K.M., Wolfe, A.M. \& Turnshek, D.A. 1993, preprint

Lu, L., Wolfe, A.M. \& Turnshek, D.A. 1991, ApJ, 361, 19

Navarro, J.F. \& White, S.D.M. 1994, MNRAS, in press

Olive, K.A., Schramm, D.N., Stegman, G. \& Walker, T.P 1990, Phys. Lett. B,
236, 451

Pogosyan, D.Y. \& Starobinsky, A.A. 1993, MNRAS, 265, 507

Rees, M.J.R. 1986, MNRAS, 218, 25p

Scalo, J.N. 1986, Fundam. Cosmic Phys., 11, 1

White, S.D.M. \& Frenk, C.S. 1991, ApJ, 379, 52

White, S.D.M., Efstathiou, G. \& Frenk, C.S. 1993, MNRAS, 262, 1073

Wolfe, A.M., Turnshek, D.A., Smith, H.E. \& Cohen, R.D. 1986, ApJS, 61, 249

Wolfe, A.M., Turnshek, D.A., Lanzetta, A.M. \& Oke, J.B. 1992, ApJ, 385, 151

Wolfe, A.M., Turnshek, D.A., Lanzetta, K.H. \& Lu, L. 1993, ApJ, 385, 151

\pagebreak

\Large
\begin {center} {\bf Figure Captions} \\
\end {center}
\vspace {1.5cm}
\normalsize
\parindent 7mm
\parskip 8mm
\baselineskip 0.8 cm

{\bf Figure 1:} The evolution of the mass density of HI gas as a function of
redshift.
The points with error bars are from Lanzetta et al. (1993). The
long- dashed, dotted and short-dashed lines show results for a b=1.5 CDM, b=2.5
CDM and
MDM model respectively, with a star formation rate proportional to the cooling
rate
of gas in halos. The thick solid line shows results for a b=1.5 CDM model in
which
stars form less efficiently at high redshift. The short-dashed/dotted line
shows the
contribution from all baryons present in halos $50 < V_c < 300$ km s$^{-1}$ in
the
MDM model, and the long-dashed/dotted line shows the same thing for halos with
$V_c > 35$ km s $^{-1}$.

{\bf Figure 2:} The column density distribution of damped Ly$\alpha$ systems in
three
redshift bins. Symbols with error bars are from the data of Lanzetta et al.
(1993) and lines
show the predictions of the model described in section 4.

{\bf Figure 3:} ({\em a}) The star formation rate as a function of redshift for
the central galaxy
of a dark matter halo with $V_c=220$ km s$^{-1}$.
({\em b}) Comparison of the predicted spectrum at $z=0$
({\em thick lines}) with the observed
spectra of two typical nearby Sb and Sc galaxies ({\em thin lines}).
The model spectrum was computed using the Bruzual \& Charlot
(1993) spectral synthesis code and has a Scalo IMF with lower
and upper cutoffs of $0.1\,M_\odot$ and $100\,M_\odot$, respectively.

\pagebreak
\end {document}